\documentclass[twocolumn,floatfix,prb,eqsecnum,showpacs,a4paper]{revtex4}
\usepackage{graphicx}
\usepackage{amsmath}
\usepackage{bm}

\begin{document}

\title{Linear response of a one-dimensional conductor coupled to a dynamical impurity with a Fermi edge singularity}

\author{I. Snyman}
\email{izaksnyman1@gmail.com}
\affiliation{School of Physics, University of the Witwatersrand, PO Box Wits, Johannesburg, South Africa}
\date{December 2013}
\begin{abstract} 
I study the dynamical correlations that a quantum impurity induces in the Fermi sea to which it is coupled.
I consider a quantum transport set-up in which the impurity can be realised in a double quantum dot. The same Hamiltonian 
describes tunnelling states in metallic glasses, and can be mapped onto the Ohmic spin-boson model. It exhibits a Fermi edge singularity,
i.e. many fermion correlations result in an impurity decay rate with a non-trivial power law energy dependence.
I show that there is a simple relation between temporal impurity correlations on the one hand, and the linear response
of the Fermi sea to external perturbations on the other. This results in a power law singularity 
in the space and time dependence of the non-local polarisability of the Fermi sea, which can be detected in transport experiments. 
\end{abstract}
\pacs{73.40.Gk, 72.10.Fk}
\maketitle

\section{Introduction}

Often, when a Fermi sea couples to the localised degree of freedom of a quantum impurity, the dynamics and
thermodynamics of the impurity are non-trivially affected.\cite{hew, Wei93, mou, naz09} 
In turn, the impurity induces correlations between the otherwise non-interacting electrons in the Fermi sea.\cite{cos, oh, hol, aff, mit, med} 
At present, we have a more complete understanding of the dissipative dynamics of the impurity than we have of
impurity induced spatial and temporal correlations induced in the Fermi sea.\cite{mit} 
Motivated by this relative lack of understanding, I
study a quantum transport setup where an impurity interacts with electrons in a one-dimensional conductor.
The internal dynamics of the impurity is restricted to a two-dimensional Hilbert space. In the past, 
the model has been used to account for the low temperature thermodynamics
of metallic glasses, in terms of tunnelling atoms coupled to the conduction electrons.\cite{kon} 
However, as in the case of the Kondo effect,\cite{pus04} a realisation of the model
in a quantum transport setup allows for greater tunability,\cite{aba04,hey} and a wider variety of possible measurements.
Whereas the electron-electron correlations that I study would be hard to detect in a metallic glass,
they are imminently observable in the setup I study.

The following is known about the system.
In the weak tunnelling limit, the
exponential decay rate $W(\varepsilon)$ associated with the relaxation of the impurity has a
power law singularity\cite{sny07,sny08} $\sim \varepsilon^{2\alpha-1}$. Here $\varepsilon$ is the energy bias between the two
impurity states and $\alpha$ is the coupling strength between the impurity and the Fermi sea. 
The power law form of $W(\varepsilon)$ is known as a Fermi edge singularity.\cite{mah67,noz69,oth90} 
It is a non-trivial many body effect involving a sum to infinite order of an expansion
in $\alpha$. In this expansion, higher powers imply more particle-hole excitations in the conductor.
The involvement of this multitude of particle-hole excitations in the impurity decay process is known as Fermi sea shake-up.\cite{ada}
The results that I report in this article reveal a simple relation between charge fluctuations associated with Fermi sea shake-up
and the decay rate $W(\varepsilon)$.

Further insight into the dynamics of the impurity is obtained by bosonising the Fermi sea.\cite{Hal81,del98} This maps 
the system onto the Ohmic spin-boson model.\cite{gui,Wei90,She,Sny} (Unlike the mapping between the spin-boson model and
the anisotropic Kondo model, here the mapping is exact.) 
It reveals that the impurity undergoes a localisation-delocalisation
quantum phase transition at $\alpha=1$.\cite{Leg87}  
The analysis leading to this conclusion is typical of many studies into open quantum systems, in that the bath degrees
of freedom are traced out at an early stage.\cite{Fey}
This is the appropriate approach for addressing fundamental questions regarding dissipation and decoherence in quantum mechanics. 

In this article I take a different perspective. 
I study the correlations that the two level system induces between the otherwise non-interacting degrees of freedom of the bath.
This is in the same spirit as recent studies on the screening cloud around an impurity that displays a Kondo effect.\cite{cos, oh, hol, aff, mit, med}
In a previous work, I considered static density correlations among electrons in the system's ground state.\cite{Sny} (Static here means correlations between
densities at equal times but at different points in space.) Long range correlations were found. Thanks to the Fermi edge singularity, these
correlations have a power law dependence on $\varepsilon$, the power law exponent being $2\alpha-3$. In the present work, I take the
logical next step and investigate dynamic density correlations. I also generalise to arbitrary temperatures. 
It is important to ask whether the electron-electron correlations that I study are observable. In many open systems, bath degrees
of freedom are not directly accessible to outside observers, but only indirectly through their effect on the impurity.   
I will show that in the setup I consider, a striking signal is produced in the electron transport trough the conductor.

I calculate
the conductor's non-local polarisability, that measures the linear response of the electron density to a potential fluctuation.
The polarisability is affected by the interaction with the impurity as follows.
When the system is perturbed by means of a potential fluctuation, a charge fluctuation is  
generated. The charge fluctuation propagates towards the impurity. As it passes the impurity, it sets it
in motion. The original fluctuation is distorted by this excitation process but continues to propagate towards the detector. 
The excited impurity then acts back on the electrons in the conductor, creating further charge fluctuations.
These are also picked up by the detector.  
One of the main results of the present work is that the polarisability 
has a power law singularity as a function of time after first arrival of the signal at the detector. This can be traced back
to the Fermi edge singularity. The power law exponent is found to be $-2\alpha$, the same as that of the Fourier transform of $W(\varepsilon)$.

The plan for the rest of the article is as follows. In Sec.\,\ref{s2}, the Hamiltonian for the model that I study is presented. The connection to the Fermi 
edge singularity and the mapping
to the spin-boson model are explained. The questions
that will be answered in the rest of the text are formulated precisely. In Sec.\,\ref{s3}, a relation between electron and impurity correlations
is derived. The results of this section are exact. In Sec.\,\ref{s4}, an exact expression for the polarisability is derived at a special
value of the coupling, where an exact expression for the impurity Green function is known. Based on this expression, a regime is identified
where perturbation theory in the impurity tunnelling amplitude is valid. In Sec.\,\ref{s5}, the leading order term in this expansion is calculated
for arbitrary coupling. Sec.\,\ref{s6} contains a summary of results.

\section{Model}
\label{s2}

I study a setup in which electrons in a one-dimensional conductor interact with a two-level impurity. The nature of the interaction is as follows.
The impurity creates a local electrostatic potential that scatters the electrons propagating in the conductor. 
The shape of the potential, and hence the scattering matrix of the conductor, depends on the state of the impurity.\cite{sny07,sny08,She,Sny} 
When the impurity is held fixed in respectively state $\left|+\right>$ or $\left|-\right>$, the scattering matrix of the conductor is
$S_+=e^{-iU_+}$ or $S_-=e^{-iU_-}$. I denote the dimension of the scattering matrices by $M$ and refer to this matrix structure as channel space. 
Furthermore, there is a tunnelling amplitude $\Delta$ and an energy bias $\varepsilon_0$ between impurity states
$\left|+\right>$ and $\left|-\right>$. An impurity of this type can be realised by a single electron trapped in a double
quantum dot.\cite{elz03,pet04} Here $\left|\pm\right>$ would correspond to the electron being localised in the ground state of the one or the other of the two
dots. The trapped electron produces an electrostatic potential that is felt by the electrons in the conductor. This potential
depends on which one of the states $\left|+\right>$ or  $\left|-\right>$ the trapped electron occupies. 

An effective low-energy description is provided by the Hamiltonian
\begin{equation}
H=H_+P_++H_-P_-+\frac{\varepsilon_0}{2}\sigma_z+\frac{\Delta}{2}\sigma_x,
\end{equation}
with 
Pauli matrices $\sigma_x=\left|+\right>\left<-\right|+\left|-\right>\left<+\right|$ 
and $\sigma_z=\left|+\right>\left<+\right|-\left|-\right>\left<-\right|$ acting
on the impurity's internal degree of freedom, and
$P_\pm=(1\pm\sigma_z)/2$ projection operators onto impurity states $\left|\pm\right>$. The 
fermion Hamiltonians 
\begin{equation}
H_\pm=\int dx\, \bm\phi^\dagger(x)\left[-i\partial_x+U_\pm f(x)\right]\bm\phi(x),\label{eq1b}
\end{equation}
describe the electrons in the conductor when the impurity is held fixed in the state $\left|\pm\right>$.
I use units were the Fermi velocity equals unity.
In the last equation, $\bm\phi(x)$ is an $M$ dimensional column vector of fermion annihilation operators, 
the vector structure referring to channel space.
 To arrive at this form, the dispersion relation
around the Fermi energy was linearised. 
In the usual one-dimensional Fermi gas, that contains both left- and right-moving electrons, (\ref{eq1b})
is obtained by 
the standard trick of unfolding of the scattering channels, so that all propagation is from left to right.\cite{fab95}
Thus $x<0$ refers to electron amplitudes incident on the impurity, while $x>0$
refers to outgoing amplitudes. The diagonal matrix elements of $U_\pm$ describe forward scattering off the impurity
in the unfolded representation, i.e. intra-channel reflection in the physical or ``folded'' picture. The off-diagonal
elements of $U_\pm$ describe inter-channel scattering in in the unfolded representation, which
corresponds to transmission or inter-channel reflection in the physical picture. 
It is also possible to realise a Fermi gas in which there are only right mover, say. An
example is a quantum Hall edge channel. For these realisations, (\ref{eq1b}) follows without unfolding.
The function $f(x)$ is sharply peaked at $x=0$, the position of the impurity, with
$\int dx\,f(x)=1$, i.e. it is delta function like. 

The Hamiltonian $H$ results from integrating out those high energy degrees of freedom
for which the linear dispersion and the truncation of the impurity Hilbert space break down. As a result, the length scale $1/\Lambda$ on
which $f(x)$ varies, is longer than the actual scale on which the impurity interaction varies. It is at least a few times the Fermi wave length. In 
other words, the ultra-violet scale $\Lambda$ is at most a fraction of the Fermi energy. 
Below, $f(x)$ will be replaced by a delta
function. When an ultra-violet regularisation is required, the delta function will be taken as a Lorentzian
\begin{equation}
f(x)=\frac{1}{2\pi\Lambda}\frac{1}{x^2+(1/2\Lambda)^2}.
\label{eq0}
\end{equation}
I will only be concerned with physics at length scales larger than $1/\Lambda$.

The Hamiltonian can be replaced by a form that is diagonal in channel space as follows.
Define new fermion operators
\begin{equation}
\bm\varphi(x)=e^{iU_+g(x)}\bm\phi(x),
\end{equation}
with 
\begin{equation}
g(x)=\int_0^xdx'\,f(x').
\end{equation}
In terms of these operators, $H_\pm$ read
\begin{eqnarray}
H_+&=&\int dx\,\bm\varphi^\dagger(x)(-i\partial_x)\bm\varphi(x),\nonumber\\
H_-&=&\int dx\,\bm\varphi^\dagger(x)\left[-i\partial_x +\tilde U(x)f(x)\right]\bm\varphi(x),\label{eq1c}
\end{eqnarray}
with $\tilde U(x)=e^{iU_+g(x)}[U_--U_+]e^{-iU_+g(x)}$.
Now consider the single particle Schr\"odinger equation associated with $H_-$, namely
\begin{equation}
E\bm\varphi_1(x)=\left[-i\partial_x +\tilde U(x)f(x)\right]\bm\varphi_1(x),\label{eq1d}
\end{equation}
where $\bm\varphi_1(x)$ is an $M$ component single particle wave function. This equation is solved by
\begin{equation}
\bm\varphi(x)=e^{iEx}e^{iU_+g(x)}e^{-iU_-g(x)}\bm\varphi_-.
\end{equation}
Utilising the fact that $g(x)$ approaches $1$ to the right of the impurity, we identify the scattering matrix associated with (\ref{eq1d}) as $S_+^\dagger S_-$. 
For the low energy physics I am interested in, the short wavelength structure of the potential in (\ref{eq1c}) and (\ref{eq1d})
is irrelevant. I therefore replace the potential with a simpler potential that produces the same scattering matrix, namely
$Vf(x)$, where $V=i\ln S_+^\dagger S_-$.\cite{aba04} The branch of the logarithm is fixed by the requirement that $V$
should evolve continuously from zero as the overall coupling strength between the conductor and the impurity is varied from zero to full strength. 
Finally, the unitary transformation $\bm\psi(x)=Q^\dagger e^{iVg(x)/2}\bm\varphi(x)$, with
$Q$ the matrix that diagonalises $V$, leads to the form
\begin{align}
H=&\sum_l\int dx\, \psi^\dagger_l(x)\left[-i\partial_x+\gamma_l\delta(x)\sigma_z\right]\psi_l(x)\nonumber\\
&+\frac{\varepsilon}{2}\sigma_z+\frac{\Delta}{2}\sigma_x.\label{eq1}
\end{align}
Here  $\gamma_l$ is the phase shift in channel $l$ of the combined scattering matrix $S_+^\dagger S_-$,
i.e. $\gamma_l$ are the eigenvalues of $-i\ln(S_+^\dagger S_-)/2$. Also, as advertised above, $f(x)$ has been replaced with
a delta function. From here on, I will use the representation (\ref{eq1}) of the
Hamiltonian, referring to it as the representation in the $\psi$-basis, as opposed to the $\phi$-basis of (\ref{eq1b}).

In this derivation, a subtle issue was glossed over. It involves a contribution to the impurity bias and is reflected in (\ref{eq1}) by the 
replacement of $\varepsilon_0$ with $\varepsilon$. It 
has to do with the singular nature of a density operator such as $\phi^\dagger_l(x)\phi_l(x)$ for fermions with a linear dispersion,
and hence an infinitely deep Fermi sea. Naively, it would seem that for such fermions, a Hamiltonian containing a scalar potential $v(x)$
can be transformed into a free Hamiltonian, using a gauge transformation $\phi(x)=\exp{-i\int_0^{x}dx'\,v(x')}\tilde\phi(x)$, that
leaves the density unaffected. This would mean that no potential could trap any charge. Careful regularisation of the problem,
for instance by means of point splitting or bosonisation,\cite{del98} shows that this is not the case. The transformed density operator
turns out to be $\tilde\rho(x)=\rho(x)-v(x)/2\pi$, thereby accounting for the missing charge. 

The only effect of
implementing a regularised version of
the derivation of (\ref{eq1}) is to change $\varepsilon_0$ to $\varepsilon=\varepsilon_0+\mbox{ offset}$,
 with the offset depending
in the short distance details of the impurity interaction. Such a contribution to the bias is clearly required to account 
for the interaction with the ``missing'' charge after the
transformation from $\phi$ to $\psi$. From here on I treat $\varepsilon$ as a phenomenological parameter that can
be adjusted by varying the external bias between impurity states $\left|+\right>$ and $\left|-\right>$.
In Sec. \ref{s5}, where I trace out the fermions, a further contribution to $\varepsilon$, that again depends on the short distance
details of the impurity interaction, will be found. At that point, I will simply redefine $\varepsilon$ to incorporate the new contribution too,
rather than using a different symbol.  
A further point to note is that the density of $\sum_l\psi^\dagger_l(x)\psi_l(x)$
of $\psi$-fermions in (\ref{eq1}) is in general not the same as the density of $\phi$-fermions in (\ref{eq1b}). However, as indicated
in the previous paragraph, the two operators only differ where the impurity interaction, which is proportional to $f(x)$, is non-zero.
Thus, away from the impurity, the subtle issue of the difference between $\phi$- and $\psi$-densities can be ignored. 

As mentioned in the introduction, in its original fermionic incarnation, the system displays a Fermi edge singularity. In this context, the following
is relevant. Consider the regime of sufficiently large $\varepsilon$ and the system initialised with the impurity in the state $\left|+\right>$ and the 
conductor in the ground state of $H_+$. As a function of time, the expectation value $\left<\sigma_z(t)\right>$ will decay from $1$ at $t=0$
to a value of $-1+\mathcal O(\Delta_r/\varepsilon)$, where $\Delta_r$ is the effective tunnelling amplitude of the impurity. Its precise definition (\ref{dr})
is deferred for two paragraphs, until I've introduced some concepts related to the spin-boson model.
For times larger than $1/\Lambda$, the decay is exponential $\sim \exp[-W(\varepsilon)t]$.
For $\varepsilon\ll\Lambda$, the decay rate is given by\cite{She} 
\begin{equation}
W(\varepsilon)=\frac{\pi\Delta_r}{2\Gamma(2\alpha)}\left(\frac{\Delta_r}{\varepsilon}\right)^{1-2\alpha},\label{eq1f}
\end{equation}
with
\begin{equation}
\alpha=-\frac{1}{2}{\rm tr}\frac{(\ln S_+^\dagger S_-)^2}{4\pi^2}.\label{eq1bb}
\end{equation}
One of the main results I derive in Sec.\,\ref{s5} is a relation between the rate $W(\varepsilon)$ and the charge associated
with the system's response to a perturbation in the external electrostatic potential.

With the aid of bosonisation, the Hamiltonian of (\ref{eq1}) maps onto the spin-boson model with an Ohmic bath, i.e. at low frequencies,
the bath spectrum is linear.\cite{gui,Wei90} At frequencies larger than $\Lambda$, the scale set by $f(x)$, the bath spectral density falls of to zero.
The precise detail of how it does so depends on the shape (but not the overall magnitude) of $f(x)$,\cite{She} but only affects the correlations
I study at short times and distances ($<1/\Lambda$). The Lorentzian regularisation (\ref{eq0}) of $f(x)$ corresponds to
the conventional choice of a spectral density that decays exponentially at large
frequencies.\cite{Leg87} The bath spectral density is then given by
\begin{equation}
J(\omega)=2\pi\alpha\omega e^{-\omega/\Lambda}.
\end{equation}  
(A redundant coupling constant between the bath and the impurity, conventionally denoted as $q_0$ in the spin-boson model, has been set equal to $1$.)
The parameter $\alpha$ defined
in (\ref{eq1bb}) is identical to the parameter $\alpha$ of Ref. [\onlinecite{Leg87}] and $K$ of Refs. [\onlinecite{Wei93}] and [\onlinecite{Wei90}].
It characterises the coupling strength between the bath and the impurity, and hence also the dissipation strength.
As another indication of the non-trivial effect that the bath has on the impurity, I mention in passing that 
the Ohmic spin-boson model undergoes a quantum phase transition
at $\alpha=1$. For $\alpha<1$, a system that is initially prepared with the impurity in one of the states $\left|\pm\right>$ and the bath in the
corresponding ground state of $H_\pm$, eventually relaxes so that the reduced density
matrix of the impurity approaches its equilibrium form, even when $\varepsilon=0$. For $\alpha>1$ and $\varepsilon=0$ however, the
impurity never relaxes but remains stuck in the state it is initialised in. 

It is useful to define an effective impurity tunnelling amplitude
\begin{equation}
\Delta_r=\Delta\left(\frac{\Delta}{\Lambda}\right)^{\alpha/(1-\alpha)}.\label{dr}
\end{equation} 
It turns out that physical quantities that are not ultra-violet divergent (i.e. insensitive to physics at length scales $1/\Lambda$ or shorter) 
only depend on $\Delta$ and $\Lambda$ through $\Delta_r$.\cite{Leg87,Wei93} It should further be noted that $\Delta$ itself is an effective tunnelling amplitude
that emerges after the impurity Hilbert space has been truncated to two dimensions by integrating out high energy excited states. In 
Refs. [\onlinecite{Leg87}] and [\onlinecite{Wei93}] it is shown that this leads to a dependence $\Delta\sim \Lambda^\alpha$ so that
$\Delta_r$ is in fact independent of $\Lambda$. Rather, the ultra-violet scale that $\Delta_r$ is sensitive to, is the one
at which restricting the dynamics of the impurity to a two-dimensional Hilbert space breaks down. (This scale is larger than $\Lambda$).
I will treat $\Delta_r$ as a phenomenological parameter characterising the effective
impurity tunnelling amplitude, and express final answers in terms of it. 
A final fact to note about the Ohmic spin-boson model, is that it is
exactly solvable for $\alpha=1/2$.\cite{Wei90} This will allow me to calculate electronic correlation functions exactly in Sec.\,\ref{s4} for this
specific value of the dissipation parameter. Apart from being illuminating in their own right, these exact results support the
perturbative analysis that is done in Sec.\,\ref{s5} for arbitrary $\alpha$.

My aim is to study the effect that the non-trivial dynamics of the impurity has on electron transport, and in particular
on the electronic correlation functions relevant for linear response. Ideally then, I should calculate an arbitrary two-electron retarded Green function.
However, results for such a general object have thus far proven elusive. What I report here are results for a particular type of two-particle Green function,
namely, the polarisability in the $\psi$-basis
\begin{equation}
\chi_{ll'}(x,y,t)=-i\theta(t)\left<\left[\rho_l(x,t),\rho_{l'}(y,0)\right]\right>,
\label{eqchi}
\end{equation}
where $\rho_l(x)=\psi_l^\dagger(x)\psi_l(x)$ is the density at $x$ in channel $l$. The linear response at time $t$
and position $x$ of the density in channel $l$ to a potential perturbation $v=\sum_{l'}\int dx'\,v_{l'}(x',t)\rho_{l'}(x')$ is
\begin{equation}
\left<\Delta\rho_l(x,t)\right>=\sum_{l'}\int dy\int dt'\,\chi_{ll'}(x,y,t-t')v_{l'}(y,t').
\end{equation}
Since perturbing or measuring in individual channels of the $\psi$-basis does not seem realistic,
it makes sense to sum over $l$ and $l'$, thereby obtaining the total linear response $\chi(x,y,t)$ at $x$ to a perturbation at $y$.
It is important to remember here that $x$ and $y$ refer to positions in the unfolded channels. A perturbation or measurement
that is local (i.e. occurs at a single position) in the unfolded representation, is non-local, occurring simultaneously at two points,
in the folded, physical representation. This however by no means implies that $\chi(x,y,t)$ is unobservable. In a system 
containing both left- and right-movers, one simply has to perform four kinds of measurements. First, one would perturb the potential
at a distance $|y|$ to the left of the impurity, and measure the fluctuation in the densities at a distance $|x|$ both to the left and to the right 
of the impurity, taking care to subtract the component of the fluctuation that reaches the left detector before interacting with the impurity.
Then one would repeat the procedure, now with an identical potential perturbation at a distance $|y|$ to the right of the detector.
Finally one would add the four measured density fluctuations. It is this total density fluctuation that is described by the polarisability
$\chi(x,y,t)$. Note also that in realisations where all the electrons move in the same direction (such as a quantum Hall edge channel),
these considerations don't apply, and $\chi(x,y,t)$ can be obtained from a single measurement.

The impurity induced part of $\chi_{ll'}(x,y,t)$ (with $x<0$ and $y>0$)
consists of two parts, namely a delta-pulse $-Q_{ll'}\delta(y+t-x)$, followed by a decaying tail.
Different physical processes are responsible for these two contributions. The delta-pulse results from the original excitation
of the impurity, while the tail results from the subsequent interaction of the excited impurity with the conductor. The two contributions
are however not completely independent. Owing to charge conservation, $\int_{x-y+0^+}^\infty dt\, \chi_{ll'}(x,y,t)=Q_{ll'}$. 
I call $Q_{ll'}$ the response charge, and calculate it below, it being a single number that characterises the strength of the impurity
induced electron-electron response.
      
Before embarking on the analysis, I comment on the connection between the system studied here and the anisotropic Kondo model.
Using the bosonisation route, a mapping between the two has been established.\cite{gui}  
This raises the question whether the correlations that I consider, can also be observed in a system with a Kondo impurity.
The answer is negative. In parts of parameter space, there is indeed a one to one correspondence between
the dynamics and thermodynamics of the two level system in the spin-boson model, and an anisotropic Kondo
impurity.\cite{cos2} However, this correspondence does not extend to bath degrees of freedom. 
The reason is that, in contrast to the mapping I employ, 
the mapping from the Kondo model to the spin-boson model is not exact:\cite{Leg87,Wei93,kei} 
in technical terms, Klein factors that are associated with different fermion species, and hence don't cancel,
are dropped. The presence or absence of Klein factors does not seem to have 
an important effect on the dynamics of the impurity. 
However, they fundamentally change the nature of electron-electron correlations.\cite{del98}
The basic result on which the analysis presented in this article relies, namely the Dzyaloshinskii-Larkin theorem,\cite{dzy,ler}
does not hold for a Kondo-type coupling between the conductor and the impurity.
      
\section{Relation between impurity and electron-electron correlations}
\label{s3}
In this section I derive a simple relation between generating functionals for electron and for impurity correlations.
I use this to prove a linear relation between the polarisability $\chi_{ll'}(x,y,t)$ and the
retarded Green function of the impurity observable $\sigma_z$.
The results of this section are non-perturbative and exact.
  
The generating functional for imaginary time density-density correlations at inverse temperature $\beta$ is
\begin{equation}
F_\rho[V]=\ln{\rm tr}\left\{\mathcal Te^{-\int_0^\beta d\tau\,\left[H+\sum_l\int dx\,V_l(x,\tau)\rho_l(x)\right]}\right\}.
\end{equation}
Here $\mathcal T$ time-orders the exponential with the smallest time argument to the right, and $\rho_l(x)=\psi^\dagger_l(x)\psi_l(x)$
is the (Scr\"odinger picture) density operator at point $x$ in channel $l$. Functional derivatives with respect to $V$, evaluated at $V=0$, generate
density correlation functions.

The generating functional can be expressed as a path integral. The electronic degrees of freedom are represented by Grassmann fields
$\bar\psi$ and $\psi$. The impurity's internal degree of freedom can be represented by a complex field via for instance the $SU(2)$
coherent state construction.\cite{kla} This will have the effect of replacing Pauli-matrices $\sigma_x$ and $\sigma_z$ with complex scalar functions
$\sigma_x(\tau)$ and $\sigma_z(\tau)$. Our analysis does not require an explicit expression for the action of the non-interacting impurity,
or for the integration measure associated with the impurity degree of freedom, and these will therefore simply be denoted as 
respectively $S_0[\sigma]$ and $\mathcal D[\sigma]$. The path integral expression for $F_\rho[V]$ then reads
\begin{equation}
F_\rho[V]=\ln \int\mathcal D[\sigma]\,e^{-S_0[\sigma]}\left<e^{-S_1[\sigma,V]}\right>_0,\label{eq3}
\end{equation}
where
\begin{equation}
\left<\ldots\right>_0=\frac{1}{Z_0}\int \mathcal D\bar\psi\mathcal D\psi\ldots e^{-\bar\psi g^{-1}\psi},\label{eq4}
\end{equation}
and $g$ refers to the free electron Green function $g_k(i\Omega_n)=(i\Omega_n-k)^{-1}$. $\Omega_n$ is a fermionic Matsubara frequency and
I use the short-hand notation
\begin{equation}
\bar \psi g^{-1}\psi=\frac{1}{\beta}\sum_{ln}\int\frac{dk}{2\pi}\,\bar\psi_{lk}(\Omega_n)(i\Omega_n-k)\psi_{lk}(\Omega_n).\label{eq5}
\end{equation}
The functional 
\begin{equation}
S_1[\sigma,V]=\frac{1}{\beta}\sum_{lm}\int\frac{dk}{2\pi}\,h_{l-k}(-\omega_m)\rho_{lk}(\omega_m),\label{eq6}
\end{equation}
with
\begin{equation}
h_{lk}(\omega_n)=\gamma_l\,\sigma_z(\omega_m)+V_{lk}(\omega_m),\label{eq6a}
\end{equation}
contains the coupling of the electrons to the impurity and to the field $V$. In the above expression, $\omega_m$ is a bosonic Matsubara
frequency, and
\begin{equation}
\rho_{lk}(\omega_m)=\frac{1}{\beta}\sum_n\int\frac{dq}{2\pi}\bar\psi_q(\Omega_n)\psi_{q+k}(\Omega_n+\omega_m),\label{eq7}
\end{equation}
is the Grassmann representation of the $(k,\omega_n)$ component of the Fourier transform of the electron density in channel $l$.
In (\ref{eq4}), $Z_0=\int\mathcal D\bar\psi\mathcal D\psi\, e^{-\bar\psi g^{-1}\psi}$ refers to the partition function of the electrons 
in the absence of the impurity. A term $+\ln Z_0$ has been dropped from (\ref{eq3}), because it does not depend on $V$, and hence doesn't show up
in correllators calculated using $F_\rho[V]$.

The generating functional for imaginary time correllators for the impurity observable $\sigma_z$ is defined as
\begin{equation}
F_\sigma[B]=\ln{\rm tr}\left\{\mathcal Te^{-\int_0^\beta d\tau\,\left[H+B(\tau)\sigma_z\right]}\right\}.\label{eq8}
\end{equation}
Expressed as a path integral it reads
\begin{align}
&F_\sigma[B]\nonumber\\
&=\ln\int\mathcal D[\sigma]\,e^{-S_0[\sigma]}\left<e^{-S_1[\sigma,0]}\right>_0e^{\frac{1}{\beta}\sum_nB(-\omega_n)\sigma_z(\omega_n)}.\label{eq9}
\end{align}

I will now show that there exists a simple relation between $F_\sigma[B]$ and $F_\rho[V]$ and hence between impurity correlations and the density-density
correlations the impurity induces in the electron gas. The starting point is the path-integral expression (\ref{eq3}) for $F_\rho[V]$.
The factor $\left<e^{-S_1[\sigma,V]}\right>_0$ is the ratio between two Gaussian integrals and can therefore easily be evaluated.
Defining two operator kernels
\begin{align}
&h_{lkn,l'k'n'}=\delta_{ll'}h_{lk-k'}(\Omega_n-\Omega_{n'}),\nonumber\\
&g_{lkn,l'k'n'}=2\pi\beta\delta_{ll'}\delta(k-k')\delta_{nn'}g_k(i\Omega_n),\label{eq10}
\end{align}
one finds
\begin{eqnarray}
\left<e^{-S_1[\sigma,V]}\right>_0&=&\frac{\det[-g+h]}{\det[-g]}\nonumber\\
&=&\exp{\rm tr}\ln\left(1-gh\right).\label{eq11}
\end{eqnarray}
Remarkably, when the logarithm in (\ref{eq11}) is expanded in $gh$, it is found that all terms higher than second order are identically zero.
This is known as the Dzyaloshinskii-Larkin theorem.\cite{dzy,ler} Thus,
\begin{equation}
 {\rm tr}\ln\left(1-gh\right)=-{\rm tr}(gh)-\frac{1}{2}{\rm tr}(ghgh).\label{eq12}
\end{equation}
Explicitly evaluating the first order term, one finds
\begin{equation}
{\rm tr}(gh)=\sum_l h_{lk=0}(\omega_m=0)\left[\frac{1}{\beta}\sum_n\int\frac{dp}{2\pi}\,g_p(i\Omega_n)\right].\label{eq13}
\end{equation}
From the fundamental definition of the free fermionic Green function
\begin{equation}
g_p(i\Omega_n)=\int_0^\beta d\tau\, e^{i\Omega\tau}\int dx\, e^{-ikx}\left<\psi_l(x,\tau)^\dagger\psi_l(x,0)\right>_0,\label{eq14}
\end{equation}
follows that the term in square brackets in (\ref{eq13}) equals $\bar\rho$, the average density of electrons per channel in the absence of the impurity.
In the second order term in (\ref{eq12}) one of the frequency sums and one of the momentum integrals can be done explicitly, yielding
\begin{equation}
{\rm tr}(ghgh)=\frac{1}{\beta}\sum_{ln}\int dp\,\frac{p}{i\omega_n-p} \frac{h_{lp}(\omega_n)h_{l-p}(-\omega_n)}{(2\pi)^2}.\label{eq15}
\end{equation}  
The next step is to substitute the explicit expression for $h_{lp}(\omega_n)$ from (\ref{eq6}) into (\ref{eq13}) and (\ref{eq15}), and to separate
the resulting expressions into terms that only contain $\sigma_z$, terms that only contain $V$, and terms that contain both.
Putting it all back into (\ref{eq11}), one finds
\begin{align}
\left<e^{-S_1[\sigma,V]}\right>_0&=\left<e^{-S_1[\sigma,0]}\right>_0\nonumber\\
&\times\exp\left\{F_\rho^{(0)}[V]-\frac{1}{\beta}\sum_n B_V(-\omega_n)\sigma_z(\omega_n)\right\},
\end{align}
where 
\begin{align}
F_\rho^{(0)}[V]=&\sum_lV_{l0}(0)\bar\rho\nonumber\\
&+\frac{1}{2\beta}\sum_{ln}\int dp\,\frac{p}{i\omega_n-p} \frac{V_{lp}(\omega_n)V_{l-p}(-\omega)}{(2\pi)^2},\label{eq17}
\end{align}
is the generating functional for density correlations in the absence of the impurity, and
\begin{equation}
B_V(\omega_n)=\sum_l\int dp\,\frac{p}{i\omega_n-p}\frac{\gamma_l}{2\pi}\frac{V_{lp}(\omega_n)}{2\pi}.\label{eq18}
\end{equation}
Substituting this back into (\ref{eq3}) and comparing to the expression (\ref{eq8}) for $F_\sigma[B]$, one obtains the simple relation
\begin{equation}
F_\rho[V]=F_\rho^{(0)}[V]+F_{\sigma}[B_V].\label{eq19}
\end{equation}
 
Now consider the Matsubara Green functions
\begin{align}
&\mathcal G_{\rho}(l,p,l',q,i\omega_n)=-\int_0^\beta d\tau\,e^{i\omega_n\tau}\int dx\,dy\,e^{-ipq-iqy}\nonumber\\
&\times\left<\Delta\rho_l(x,\tau)\Delta\rho_{l'}(y,0)\right>,
\end{align}
and
\begin{align}
&\mathcal G_{\sigma}(i\omega_n)=-\int_0^\beta d\tau\,e^{i\omega_n\tau}\left<\Delta\sigma_z(\tau)\Delta\sigma_z(0)\right>,
\label{eq20}
\end{align}
 where $\Delta\rho=\rho-\left<\rho\right>$ and similarly for $\Delta\sigma_z$.  These can be obtained from the
generating functionals $F_\rho[V]$ and $F_\sigma[B]$ by means of the appropriate functional derivatives
\begin{align}
&\mathcal G_{\rho}(l,p,l',q,i\omega_n)\nonumber\\
&=-\frac{(2\pi\beta)^2}{\beta}\frac{\delta}{\delta V_{l-p}(-\omega_n)}\frac{\delta}{\delta V_{l'-q}(\omega_n)}\left.F_\rho[V]\right|_{V=0},
\label{eq21}
\end{align}
and similarly for $G_\sigma$.
Owing to (\ref{eq19}) these two Green functions are related to each other, i.e.
\begin{align}
&\mathcal G_{\rho}(l,p,l',q,i\omega_n)=\mathcal G_{\rho}^{(0)}(l,p,l',q,i\omega_n)\nonumber\\
&~~~~~~~~~~~~~~~~~~-\frac{\gamma_l\gamma_{l'}}{(2\pi)^2}\frac{p}{i\omega_n-p}\frac{q}{i\omega_n+q}\mathcal G_{\sigma}(i\omega_n),
\label{eq22}
\end{align}
where
\begin{equation}
\mathcal G_{\rho}^0(l,p,l',q,i\omega_n)=\delta_{l,l'}\delta(p+q)\frac{p}{i\omega_n-p},
\label{eq23}
\end{equation}
is the density-density Green function in the absence of the impurity.

The polarisability $\chi_{ll'}(x,y,t)$ of (\ref{eqchi}) can be obtained by analytically continuing $\mathcal G_\rho(l,p,l',q,i\omega_n)$ to real frequencies
and Fourier transforming to space and time. Similarly, the retarded impurity Green function
\begin{equation}
G_\sigma(t)=-i\theta(t)\left<[\sigma_z(t),\sigma_z(0)]\right>.
\label{eq23a}
\end{equation}
can be calculated by analytically continuing $\mathcal G_\sigma(i\omega_n)$ to real frequencies and then Fourier
transforming to time. Thus one finds
\begin{align}
&\chi_{ll'}(x,y,t)=\chi^{(0)}_{ll'}(x,y,t)\nonumber\\
&~~~~~~~~-\frac{\gamma_l\gamma_{l'}}{(2\pi)^2}\partial_x\partial_y\left[\theta(x)\theta(-y)G_\sigma(t+y-x)\right],
\label{eq24}
\end{align}
where $\chi_{ll'}^{(0)}(x,y,t)=-\delta_{l,l'}\partial_x\left[\theta(x-y)\delta(t+y-x)\right]$ is polarisability in the absence of the impurity,
in which case a delta pulse in the potential in the incoming channels at $y$ produces a density fluctuation
$\left<\Delta \rho_l(x,t)\right>=\partial_t\delta(t+y-x)$. 
The fact that this fluctuation travels at the Fermi velocity without spreading, is due to the linear dispersion of the electrons. 
The structure $\partial_t\delta(t+y-x)$ is consistent with charge conservation: since a
potential pulse cannot create charge, we must have $\int_0^\infty dt\,\chi_{ll'}(x,y,t)=0$. 

The response measured for outgoing electrons ($x>0$) to a perturbation of the incoming electrons ($y<0$) is given by
\begin{equation}
\chi_{ll'}(x,y,t)=\chi^{(0)}_{ll'}(x,y,t)+\frac{\gamma_l\gamma_{l'}}{(2\pi)^2}G_\sigma''(t+y-x),
\label{eq24b}
\end{equation}
where $G_\sigma''\equiv\partial_t^2G_\sigma$ contains the correlations induced by the impurity.
From (\ref{eq23a}) follows that $G_\sigma''(t)$ can be written as
\begin{equation}
G_\sigma''(t)=i\theta(t)R(t)-i\delta(t)\int_0^\infty dt\, R(t),
\label{eq24cc}
\end{equation}
where $R(t)= -\partial^2_t\left<[\sigma_z(t),\sigma_z(0)]\right>$. 
The total response charge in channel $l$ due to a delta potential pulse in channel $l'$, is
\begin{equation}
Q_{ll'}=i\frac{\gamma_l\gamma_l'}{(2\pi)^2}\int_0^\infty dt\,R(t).\label{eq24c}
\end{equation}
Below I will show that $Q_{ll'}$ has an interesting behaviour as a function of the coupling constant $\alpha$.

\section{Exact expressions for $\alpha=1/2$.}
\label{s4}

In order to calculate the polarisability, one needs an expression for $G_\sigma''(t)$. For $\alpha=1/2$, an exact result is available for the
Fourier transform
\begin{equation}
R(\omega)=\int_{-\infty}^\infty dt\,e^{i\omega t}R(t),
\end{equation}  
of $R$ as defined below (\ref{eq24cc}).\cite{Wei90}
The low temperature regime is the most interesting. (The role of temperature is merely to produce exponential decay with a rate $\pi/\beta$ at large times.)
I therefore specialise to zero temperature, where one has
\begin{widetext}
\begin{equation}
R(\omega)=\frac{4\Gamma}{\pi}\frac{\omega^2}{\omega^2+\Gamma^2}\left\{\arctan\frac{2(\omega+\varepsilon)}{\Gamma}+\arctan\frac{2(\omega-\varepsilon)}{\Gamma}
+\frac{\Gamma}{2\omega}\ln\left[\frac{[\Gamma^2+4(\omega+\varepsilon)^2][\Gamma^2+4(\omega-\varepsilon)^2]}{(\Gamma^2+4\varepsilon^2)^2}\right]\right\},
\label{eq25a}
\end{equation}
\end{widetext}
with $\Gamma=\pi\Delta_r/2$. 

\begin{figure}[tbh]
\begin{center}
\includegraphics[width=\columnwidth]{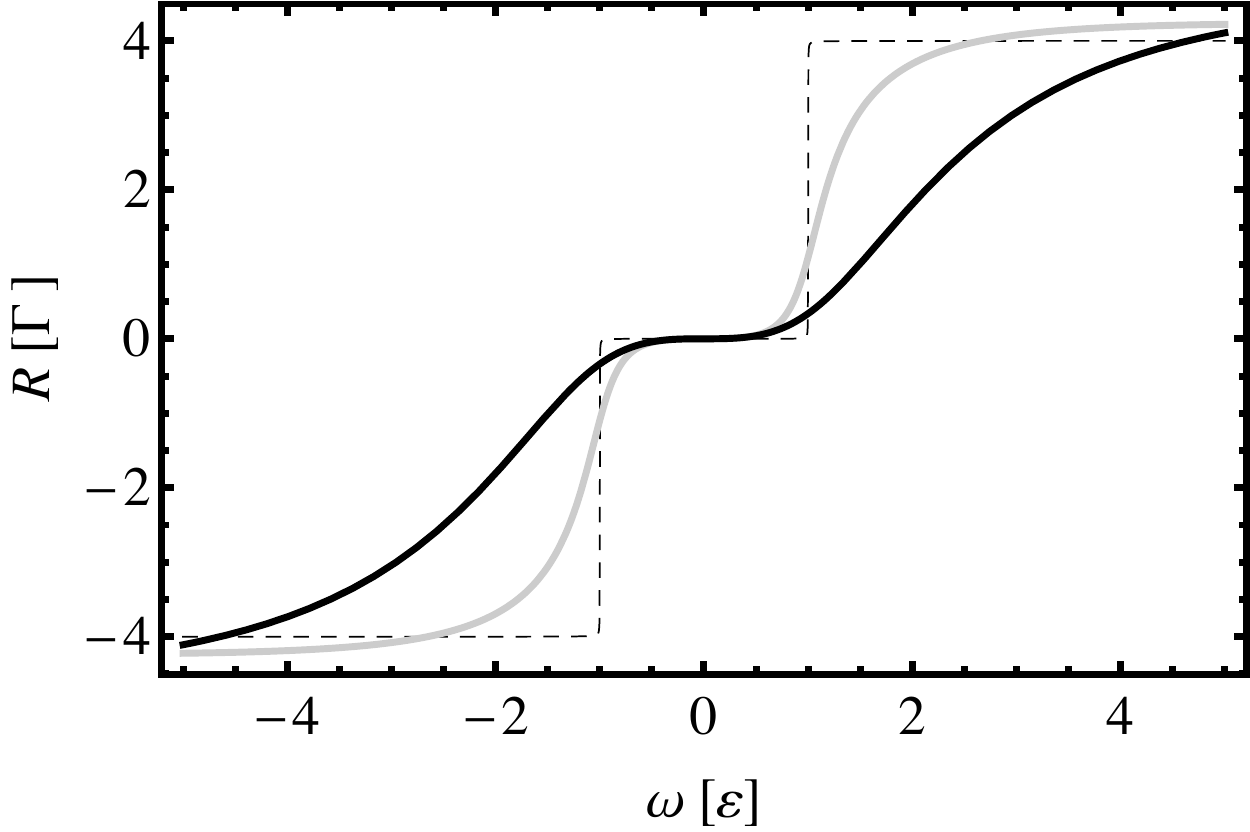}
\caption{The function $R(\omega)$, for different values of $\Gamma$. The black curve corresponds to $\Gamma=2\varepsilon$, the grey curve to $\Gamma=\varepsilon/2$
and the thin dashed curve to the limiting case $\Gamma/\varepsilon\to 0$.
\label{f2}}
\end{center}
\end{figure}
Figure \ref{f2} shows the function $R(\omega)$ for different values of $\Gamma$. As $\omega$ tends to $\pm\infty$, $R(\omega)$ tends to $\pm 4\Gamma$. For $\Gamma\to0$,
$R(\omega)/\Gamma$ is piecewise constant, with a step from $-4$ to $0$ at $\omega=-|\varepsilon|$ and a step from $0$ to $4$ at $\omega=|\varepsilon|$. As $\Gamma$ is
increased, these steps become smoothed out. 

I have not been able to perform the inverse Fourier transform that converts $R$ into $G_\sigma''$ analytically. Extracting
the small and large time asymptotics of $G_\sigma''$ is however straight-forward. The small time behaviour of $G_\sigma''$ is determined by the large
frequency behaviour of $R(\omega)$. From (\ref{eq25a}) it follows that 
\begin{equation}
G_\sigma''(t\to 0^+)\simeq\frac{4\Gamma}{\pi t}.\label{eq25b}
\end{equation}
The $1/t$ divergence in $G_\sigma''(t)$ implies that at $\alpha=1/2$, the response charge $Q_{ll'}$ suffers from a logarithmic ultraviolet divergence.
A damping factor that kicks in when $|\omega|>\Lambda$, and that is omitted from (\ref{eq25a}), regularises the $1/t$ singularity in (\ref{eq25b}) and the logarithmic
divergence in $Q_{ll'}$ at the scale of $1/\Lambda$.   

The long time behaviour of $G_\sigma''$ is determined by the analyticity structure of $R(\omega)$ in the complex $\omega$ plain.\cite{Ree}
The singularities in $R(\omega)$ at $\omega=\pm\varepsilon\pm i\Gamma/2$ and the analyticity and boundedness of $R(\omega)$ for $|{\rm Im}(\omega)|<\Gamma/2$
implies that for $t\gg\Gamma$,
\begin{equation}
G_\sigma''(t)\sim e^{-\Gamma t/2}.
\label{eq25c}
\end{equation}

\begin{figure}[tbh]
\begin{center}
\includegraphics[width=\columnwidth]{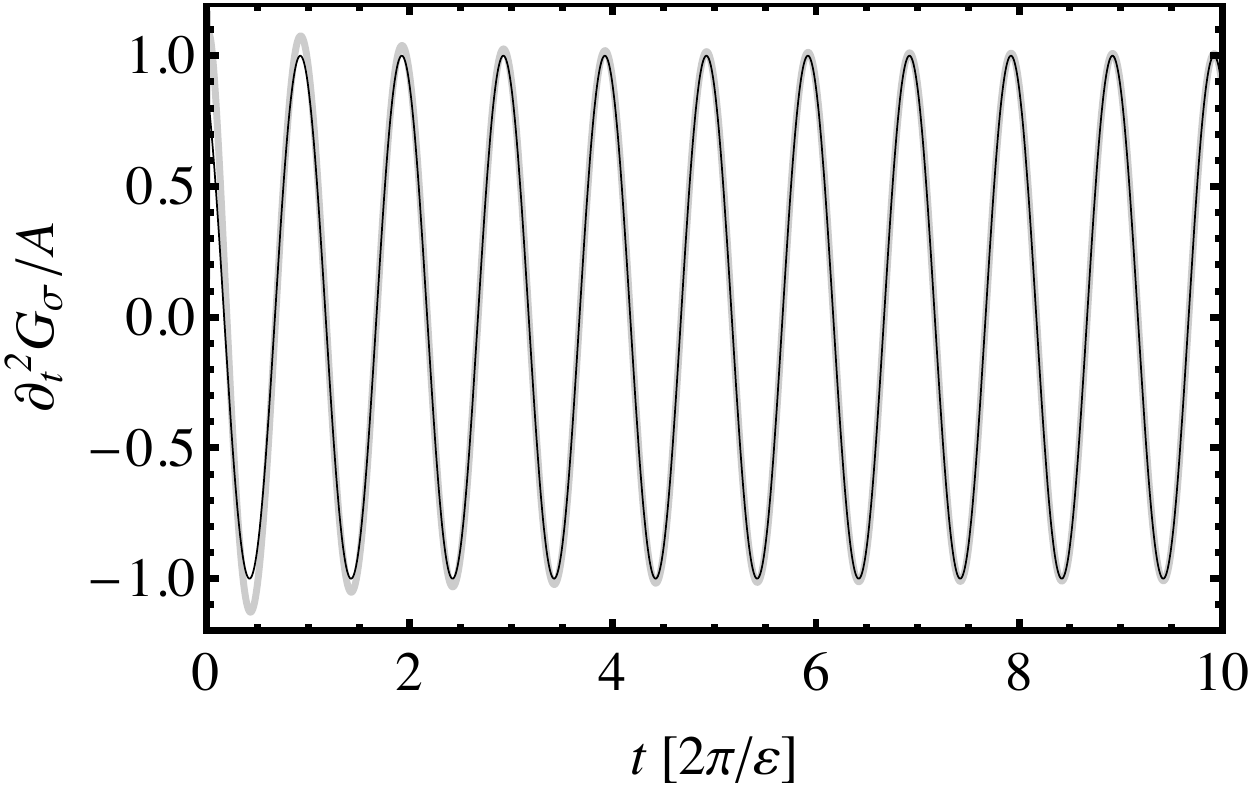}

\vspace{3mm}

\includegraphics[width=\columnwidth]{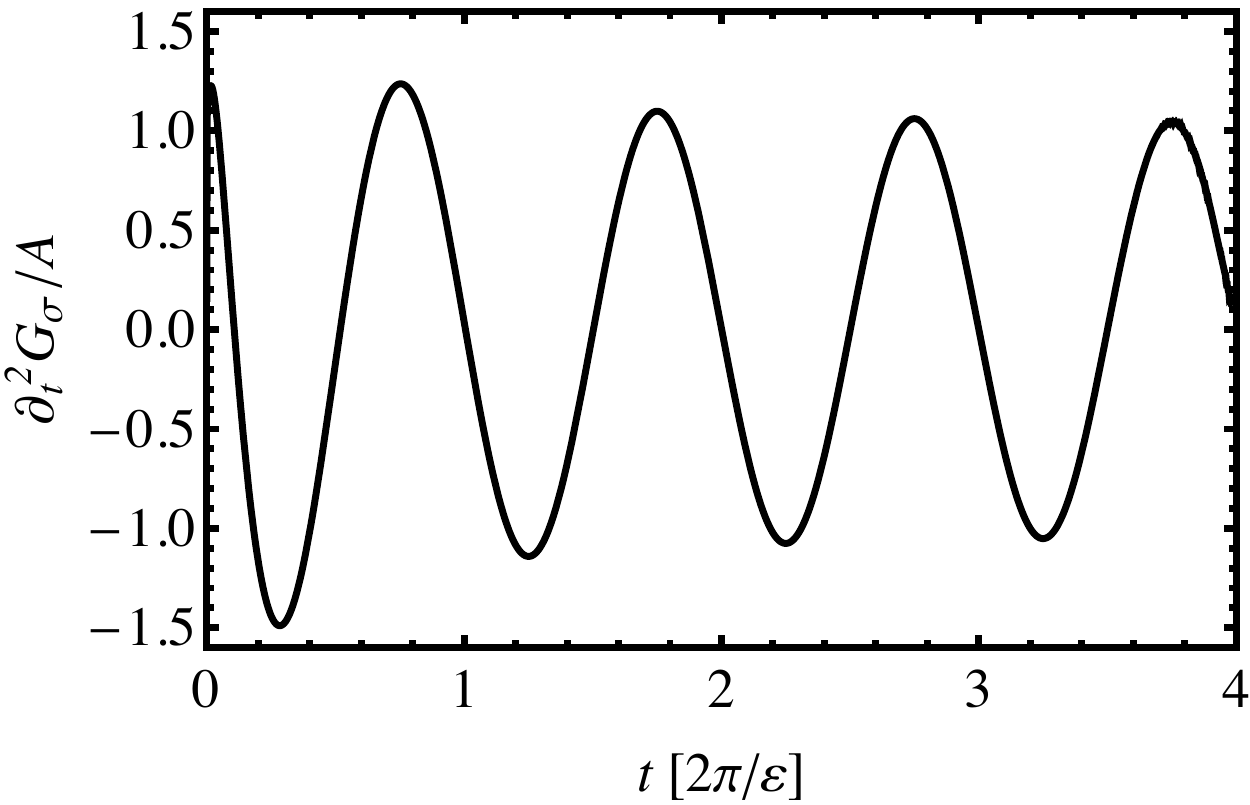}
\caption{The function $G_\sigma''(t)/A(t)$ for different values of $\Gamma$. Top: $\Gamma=\varepsilon/2$. The thin black line shows the approximation
$\cos(\varepsilon t + \Gamma/\varepsilon)$. In the time interval shown, the envelope function $A(t)$ decreases by $9$ orders of magnitude. Bottom: $\Gamma=2\varepsilon$. 
In the time interval shown, the envelope function $A(t)$ decreases by $13$ orders of magnitude. 
In both cases the first turning point, close to $t=0$, is somewhat sharper than the rest, and hence
the behaviour of the function at $t\ll\varepsilon$ is poorly resolved. Close inspection of the data (not shown) confirms that $\lim_{t\to0^+}G_\sigma''(t)/A(t)=1$
as expected.  
\label{f3}}
\end{center}
\end{figure}
By numerically performing the Fourier transform, I have found that 
\begin{equation}
A(t)=\frac{4\Gamma}{\pi t}e^{-\Gamma t/2},
\end{equation}      
excellently describes the envelope of $G_\sigma''(t)$, also for intermediate times. This can be seen in Figure \ref{f3}. In the figure, $G_\sigma''(t)/A(t)$,
with $G_\sigma''(t)$ numerically calculated from (\ref{eq25a}), is plotted for two different values of $\Gamma$. It is seen that $G_\sigma''(t)/A(t)$ oscillates
with an amplitude approaching $1$ at times larger than a few times $2\pi/\varepsilon$. I have found that, for $\Gamma<\varepsilon$, and $t>8 \pi\varepsilon$,
$G_\sigma''(t)/A(t)$ equals $\cos(\varepsilon t+\gamma/\varepsilon)$ with an error less than $1\%$. 

\begin{figure}[tbh]
\begin{center}
\includegraphics[width=\columnwidth]{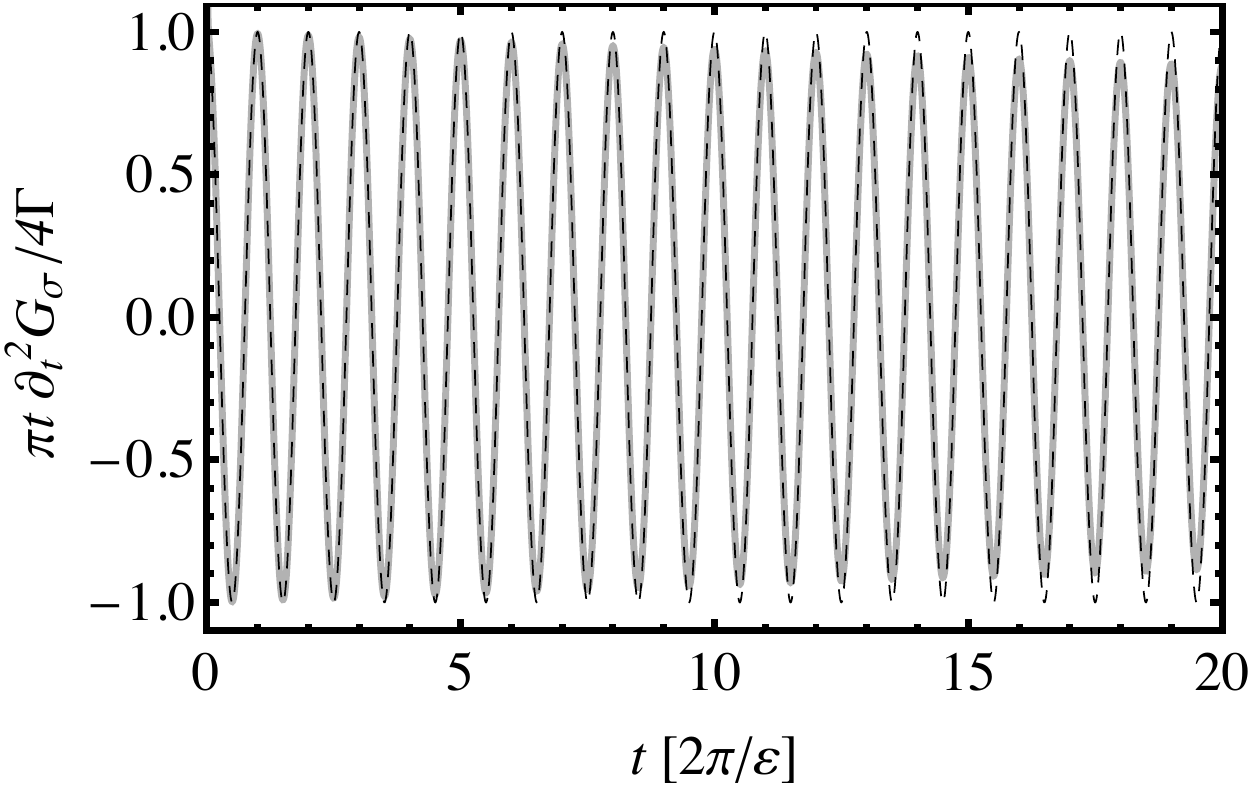}
\caption{The function $\pi t G_\sigma''(t)/4\Gamma$ for $\Gamma/\varepsilon=2\times 10^{-3}$ (solid line), compared to the $\Gamma\ll\varepsilon,\,1/t$ limiting case
$\pi t G_\sigma''(t)/4\Gamma=\cos(|\varepsilon|t)$ (dashed line) of (\ref{eq25d}). 
\label{f4}}
\end{center}
\end{figure}
Expanding $R(\omega)$ in $\Gamma$ and then performing the Fourier transform term by term, one obtains
\begin{equation}
G_\sigma''(t)=\frac{4\Gamma}{\pi}\frac{\cos|\varepsilon|t}{t} + \mathcal O(\Gamma^2).\label{eq25d}
\end{equation}        
In view of the asymptotics derived above, the status of the above expression is clear: The leading order term in the  $\Gamma\propto\Delta^2$ expansion provides an
accurate approximation to $G_\sigma''(t)$ when $\Gamma\ll \varepsilon$ for times $t\ll 1/\Gamma$. It correctly captures the oscillatory behaviour and power law envelope
that governs $G_\sigma''(t)$ for $1/\Lambda<t<1/\Gamma$, but not the eventual exponential decay at large times. The accuracy of (\ref{eq25d}) for  
$\Gamma\ll\varepsilon$ and $t\ll 1/\Gamma$ is confirmed in Figure \ref{f4}, where (\ref{eq25d}) is compared to $G_\sigma''(t)$, numerically obtained from (\ref{eq25a}),
for $\Gamma/\varepsilon=2\times10^{-3}$.

\section{Perturbative analysis at arbitrary $\alpha$.}
\label{s5}
For $\alpha\not=1/2$ no exact solution is available. In this section I therefore calculate 
$G_\sigma''(t)=-i\partial_t^2\left\{\theta(t)\left<[\sigma_z(t),\sigma_z(0)]\right>\right\}$
and hence $\chi_{ll'}(x,y,t)$, to second order in the impurity tunnelling amplitude $\Delta$.
This is the usual limit in which the Ferm edge singularity is considered.
Based on the results of the previous section, I expect the expansion to be accurate for
$\Delta_r$ sufficiently smaller than $\varepsilon$ and $t<1/\Delta_r$.
Here the expectation value is with respect to the thermal density matrix
$\exp(-\beta H)/{\rm tr}\,\exp(-\beta H)$ at inverse temperature
$\beta$ and $\sigma_z(t)=\exp(iHt)\sigma_z\exp(-iHt)$.
Expanding the operators $\exp(-\beta H)$ and $\exp(\pm i H t)$ in $\Delta$, using the interaction picture, and tracing out the impurity
degree of freedom, one finds
\begin{equation}
G_\sigma(t)=-\Delta^2\theta(t)\int_0^\beta d\tau\int_0^tdt'\,P(\tau-i t'),
\label{eq26}
\end{equation}
where
\begin{align}
&P(z)=\frac{\zeta(z)+\zeta(\beta-z)}{\zeta(0)+\zeta(\beta)},\nonumber\\
&\zeta(z)=e^{(\beta/2-z)\varepsilon}\frac{{\rm tr}\left[e^{-(\beta-z)H_-}e^{-zH_+}\right]}{Z_0},\nonumber\\
\label{eq27}
\end{align}
and 
$H_\pm$ are defined in (\ref{eq1c}). 
In order to calculate the polarisability (\ref{eq24b}), we need the second time derivative of $G_\sigma(t)$, which, from (\ref{eq26})
is given by
\begin{align}
&G_\sigma''(t)\nonumber\\
&=-\Delta^2\left\{-i\theta(t)\left[P(\beta-it)-P(-it)\right]+\delta(t)\int_0^\beta d\tau\, P(\tau)\right\}.
\label{eq28a}
\end{align}
The analyticity structure of $P(\tau)$ allows us to replace the integral from $0$ to $\beta$ in the second term by two integrals,
one along the imaginary axis from $0$ to $i\infty$ and the other along the line $\beta+i t'$ with $t'$ from $\infty$ to $0$.
From the definition (\ref{eq27}) of $P(t)$ it is seen that $P(\beta-it)=P(it)=P(-it)^*$, so that (\ref{eq28a}) becomes
\begin{equation}
G_\sigma''(t)\nonumber\\
=-2\Delta^2\left[\theta(t){\rm Im}\,P(it)-\delta(t)\int_0^\infty dt\, {\rm Im}\,P(it)\right].
\label{eq28b}
\end{equation}

Next, I calculate $\zeta(\tau)$ at real $\tau\in(0,\beta)$ and then analytically continue to complex arguments. For $\tau\in(0,\beta)$, $\zeta(\tau)$ can be written as a path integral
\begin{equation}
\zeta(\tau)=e^{\left(\frac{\beta}{2}-\tau\right)\varepsilon}\left<e^{-S_1[\sigma,0]}\right>_0,
\label{eq29}
\end{equation}
in the notation of Sect. \ref{s2}, with
\begin{align}
\sigma_{lk}(\omega_n)&=\gamma_l\int_0^\beta d\tau'\,{\rm sign}(\tau-\tau')e^{i\omega_n\tau'}\nonumber\\
&=2\gamma_l\left[\delta_{n,0}\left(\tau-\frac{\beta}{2}\right)+(1-\delta_{n,0})\frac{e^{i\omega_n\tau}-1}{i\omega_n}\right].
\label{eq30}
\end{align}
Thanks again to the Dzyaloshinskii-Larkin theorem, the path integral evaluates to a Gaussian functional in $\sigma$,
\begin{align}
&\ln\zeta(\tau)=\tilde{c}_0+\tilde{c}_1\tau-\frac{1}{2}\int dp\,\frac{1}{\beta}\sum_{ln}\frac{p}{i\omega_n-p}\left|\frac{\sigma_{lp}(\omega_n)}{2\pi}\right|^2\nonumber\\
&=c_0+c_1\tau\nonumber\\
&-\sum_l\left(\frac{\gamma_l}{\pi}\right)^2\int dp\,\underbrace{\frac{1}{\beta}\left[\sum_{n}\frac{p}{i\omega_n-p}\frac{1-\cos\omega_n\tau}{\omega_n^2}-\frac{\tau^2}{2}\right]}_{=\mathcal A}.
\label{eq31}
\end{align}
We will not require explicit expressions for the constants $c_0$ or $c_1$, which depend on microscopic detail at the scale of $1/\Lambda$.
Using the explicit expression (\ref{eq30}) for $\sigma_{lk}(\omega_n)$, the frequency sum can be evaluated by
converting it into a contour integral. Each term in the sum is associated with a pole along the imaginary axis at $i\omega_n$, $n\not=0$.
\begin{align}
&\mathcal A=-\frac{\tau^2}{2\beta}\nonumber\\
&+\frac{1}{4\pi i}\int_C dz\, \frac{1-e^{z\tau}}{2z^2}\left(\frac{p}{z-p}-\frac{p}{z+p}\right)\left(\coth\frac{\beta z}{2}-1\right).
\label{eq32}
\end{align}

\begin{figure}[tbh]
\begin{center}
\includegraphics[width=.7\columnwidth]{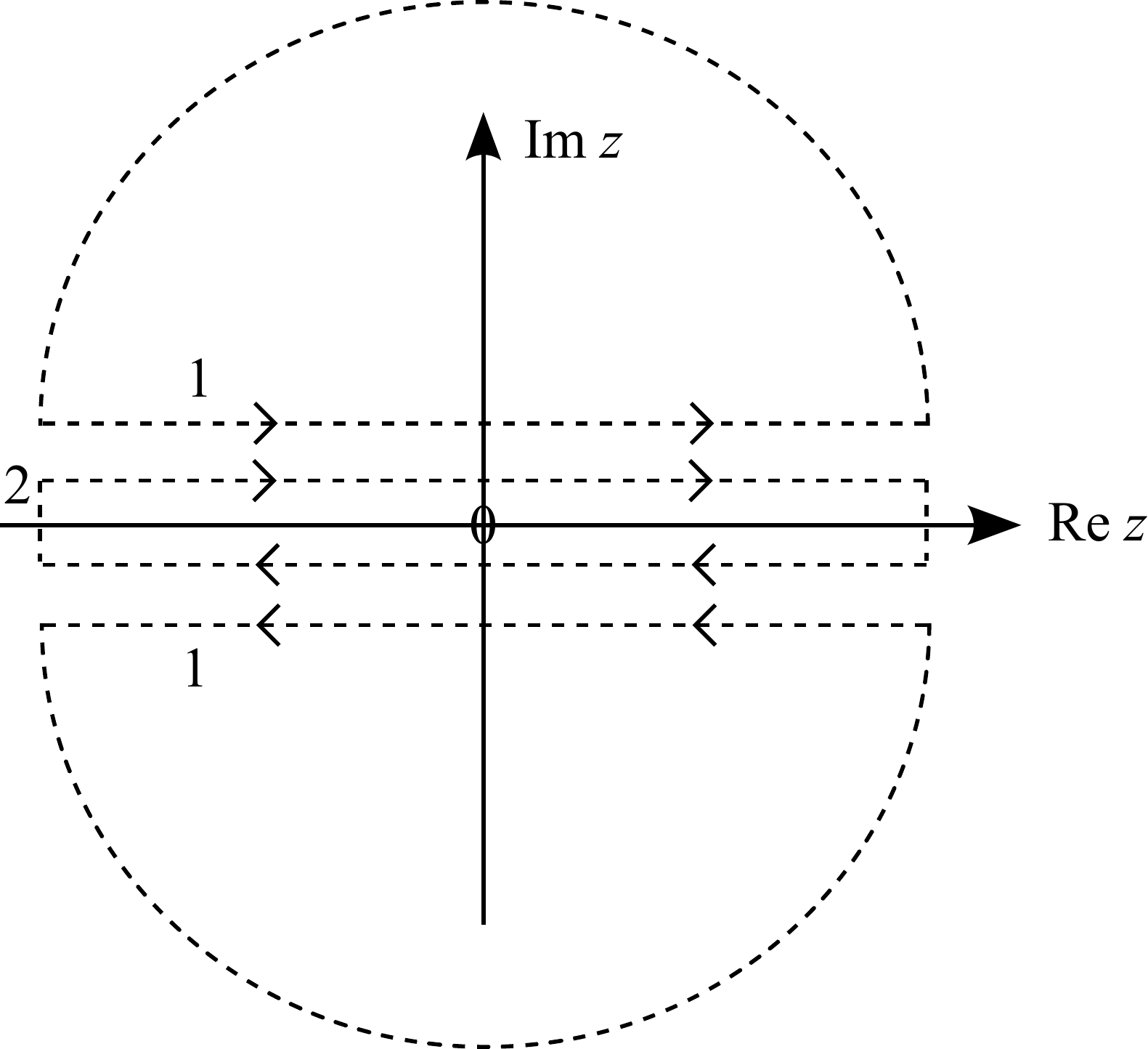}
\caption{Integration contours for evaluating the frequency sum in (\ref{eq31}).
\label{f1}}
\end{center}
\end{figure}

The contour can be deformed from contour 1 in Figure \ref{f1} to contour 2, which contains poles at 0 and $\pm p$ along the real line.
With the aid of this deformation, one finds
\begin{align}
\mathcal A=-\frac{\tau}{2}+\Bigg[\frac{1-e^{p\tau}}{4p}&\left(\coth\frac{\beta p}{2}-1\right)\nonumber\\
&+\frac{1-e^{-p\tau}}{4p}\left(\coth\frac{\beta p}{2}+1\right)\Bigg].
\label{eq33}
\end{align}
The $p$ integral in (\ref{eq31}) formally suffers from an ultra-violet divergence because I used a delta function for the impurity potential. 
Using the Lorentzian regularisation of (\ref{eq0}) and integrating
(\ref{eq33}) over $p$, leads to
\begin{align}
&-\int_{-\infty}^\infty dp\,e^{-|p|/\Lambda}\Bigg[\frac{1-e^{p\tau}}{4p}\left(\coth\frac{\beta p}{2}-1\right)\nonumber\\
&~~~~~~~~~~~~~~~~~~~~~~~~~~~~~~~+\frac{1-e^{-p\tau}}{4p}\left(\coth\frac{\beta p}{2}+1\right)\Bigg]\nonumber\\
&=\int_0^\infty\frac{dp}{p}e^{-\frac{p}{\Lambda}}\left[\frac{e^{\tau p}-1}{e^{\beta p}-1}+\frac{e^{-\tau p}-1}{1-e^{-\beta p}}\right]\nonumber\\
&=\int_{\frac{1}{\Lambda\beta}}^\infty du\int_0^\infty dv\,e^{-uv}\left[\frac{e^{\tau v/\beta}-1}{e^v-1}+\frac{e^{-\tau v/\beta}-1}{1-e^{-v}}\right] 
\label{eq34}
\end{align}
The integrals involved in the last line of (\ref{eq34}) can be done with the help of the identity
\begin{equation}
\int_0^\infty dx\, \frac{e^{px}-e^{qx}}{e^{x}-1}=\psi(1-q)-\psi(1-p),\label{eq35}
\end{equation}
with $\psi(x)=\partial_x\ln\Gamma(x)$.
The $v$ integral, evaluated at the upper boundary, vanishes so that one obtains
\begin{equation}
\zeta(\tau)=\zeta(0)e^{-\varepsilon\tau}\left[\frac{\Gamma\left(1+\frac{1}{\Lambda\beta}-\frac{\tau}{\beta}\right)\Gamma\left(\frac{1}{\Lambda\beta}+\frac{\tau}{\beta}\right)}
{\Gamma\left(1+\frac{1}{\Lambda\beta}\right)\Gamma\left(\frac{1}{\Lambda\beta}\right)}\right]^{2\alpha},\label{eq36}
\end{equation}
with $\alpha$ as defined in (\ref{eq1bb}).
 Here I have incorporated all the linear in $\tau$ terms into another redefinition $\varepsilon\to\varepsilon-\mbox{ offset}$ 
 of the impurity bias energy. Substitution into (\ref{eq27}) gives
\begin{align}
&P(\tau)\nonumber\\
&=\frac{\cosh\left(\frac{\varepsilon\beta}{2}-\varepsilon\tau\right)}{\cosh\left(\frac{\varepsilon\beta}{2}\right)}\left[\frac{\Gamma\left(1+\frac{1}{\Lambda\beta}-\frac{\tau}{\beta}\right)\Gamma\left(\frac{1}{\Lambda\beta}+\frac{\tau}{\beta}\right)}
{\Gamma\left(1+\frac{1}{\Lambda\beta}\right)\Gamma\left(\frac{1}{\Lambda\beta}\right)}\right]^{2\alpha}.\label{eq38}
\end{align}
 
Now analytical continuation of $P(\tau)$ to complex arguments is straight-forward. Further simplification is possible
if one uses the identity $\Gamma(1+z)=z\Gamma(z)$ to make the arguments of all $\Gamma$-functions non-zero in the $\Lambda\to\infty$ limit.
This allows one to take $\Lambda\to\infty$ in the arguments of the $\Gamma$-functions. Finally I employ the identity
$\Gamma(1+z)\Gamma(1-z)=\pi z/\sin\pi z$, and substitute the result into (\ref{eq28b}) to obtain for $t>x-y$ my main result
\begin{align}
&\chi_{ll'}(x,y,t)=\frac{\gamma_l\gamma_{l'}}{(2\pi)^2}G_\sigma''(t+y-x)\nonumber\\
&G_\sigma''(t)=-2\Delta_r^2\nonumber\\
&\times{\rm Im}\left[\,\frac{\cosh\left(\frac{\varepsilon\beta}{2}-i\varepsilon t\right)}{\cosh\left(\frac{\varepsilon\beta}{2}\right)}\left[
\left(\frac{1}{\Lambda}+it\right)\frac{\beta\Delta_r}{\pi t}\sinh\frac{\pi t}{\beta}\right]^{-2\alpha}\right].\label{eq39}
\end{align}
For non-zero temperatures, the result of (\ref{eq39}) implies exponential damping at a rate $2\pi\alpha/\beta$ for times larger than $\beta$. 
The $\cosh(\varepsilon\beta/2-i\varepsilon t)$ pre-factor implies damped coherent oscillations with angular frequency $\varepsilon$. 
Regardless of temperature,
the short time behaviour is a power law $\sim t^{-2\alpha}$, regularised at the scale $1/\Lambda$. This implies that
for $\alpha<1/2$ (weak damping), the response charge $Q_{ll'}$ is finite, while for $\alpha>1/2$ it diverges in the $\Lambda\to \infty$ limit
as $(\Lambda/\Delta_r)^{2\alpha-1}$. 
At the
critical point $\alpha=1/2$ it diverges logarithmically, as was also obtained from the exact solution in this case.  
The logarithmic divergence of the exact solution at $\alpha=1/2$ suggests that the divergence
found in the perturbative solution for $\alpha\geq 1/2$ is real, and not due to a breakdown in perturbation theory.

At zero temperature (\ref{eq39}) reduces to
\begin{equation}
G_\sigma''(t)\nonumber
=-2\Delta_r^2{\rm Im}\left[e^{-i|\varepsilon|t}
\left(\frac{\Delta_r}{\Lambda}+i\Delta_r t\right)^{-2\alpha}\right].\label{eq40}
\end{equation} 
At $\alpha=1/2$ and times $t\gg1/\Lambda$, this agrees with the $\Delta_r\ll\varepsilon,\,1/t$ limit of the exact $\alpha=1/2$ solution, as it should.
As was found in the case of $\alpha=1/2$, I expect (\ref{eq40}) to break down when $t$ becomes comparable to or larger than $1/\Delta_r$.
From (\ref{eq40}) the response charge in channel $l$ due to a perturbation in channel $l'$ at zero temperature and $\alpha<1/2$ is
\begin{eqnarray}
Q_{ll'}&=&2\Gamma(1-2\alpha)\frac{\gamma_l\gamma_{l'}}{(2\pi)^2}\Delta_r\left(\frac{\Delta_r}{\varepsilon}\right)^{1-2\alpha}\nonumber\\ 
&=&\frac{\gamma_l\gamma_{l'}}{\pi^2}{\rm cosec}(2\pi\alpha)W(\varepsilon),\label{eq41}
\end{eqnarray}
where $W(\varepsilon)$ is the impurity transition rate, known from the theory of the Fermi edge singularity.

\section{Summary}
\label{s6}
I studied a fermionic realisation of the Ohmic spin-boson model. The system consists of a two level impurity coupled to a one-dimensional
conductor. My aim was to characterise the dynamical correlations that the impurity induces between electrons in the conductor.  
The technical result from which the rest
follows, is a simple relation (\ref{eq19}), between a generating functional for electron-electron correlations and one for impurity correlations.  
This relation implies that for $x>0$ and $y<0$, in the unfolded coordinates (cf. Sec. \ref{s2}), the impurity contribution to the
electronic polarisability $\chi_{ll'}(x,y,t)$ defined in (\ref{eqchi}) is (cf. Eq. \ref{eq24b})
\begin{equation}
\frac{\gamma_l\gamma_{l'}}{(2\pi)^2}G''_\sigma(t+y-x),
\end{equation}
where $G_\sigma(t)$ is the retarded Green function for the impurity operator $\sigma_z$. Its second derivative
consists of a delta spike at $t=0$ followed by a decaying tail (cf Eq. \ref{eq24c}). Due to charge conservation, 
the area under the tail equals
minus the weight of the delta spike. The polarisation response of the conductor is as follows.
A potential perturbation $v(x,t)=\delta(x-y)\delta(t)$ in channel $l'$ applied to electrons incident on the impurity ($y<0$) 
produces a charge fluctuation $\partial_t\delta(t+y-x)$ among the incident electrons in channel $l'$.
This fluctuation is modified when it reaches the impurity at time $t=-y$. The modification $-Q_{ll'}\delta(x-t-y)$ in channel $l$
is due to the excitation of the impurity by the incident charge fluctuation. Behind it follows an oppositely charged decaying tail, 
produced by the subsequent interaction between the
excited impurity and the electrons in the conductor. The response charge $Q_{ll'}$ is a measure of the strength of the
impurity induced correlations in the conductor. 

For a dissipation strength $\alpha=1/2$, the available exact solution of the spin-boson model provides an exact
expression (\ref{eq25a}) for the Fourier transform of $G''_\sigma(t)$. From this I extracted the following behaviour 
of $\chi_{ll'}(x,y,t)$ (cf. Eqs. \ref{eq25b} and \ref{eq25c}). 
Coherent damped oscillations are observed. In the weak tunnelling limit, the angular frequency of these oscillations
is $\varepsilon$. 
For large times, $\chi$ decays as $e^{-\Gamma t/2}$, where $\Gamma=\pi\Delta_r/t$. For
$x-y<t\ll1/\Gamma$ it behaves as a power law $(t+y-x)^{-1}$. Due to the divergence at $t\to x-y$,
the response charge $Q_{ll'}$ is ultraviolet divergent.  
I also showed that, for $t\ll 1/\Gamma$, and $\Gamma\ll\varepsilon$, accurate results
are obtained by expanding the exact result to leading order in $\Delta$ or equivalently $\Delta_r$ (cf. Figure \ref{f4}).

I used this last insight to investigate the polarisability response of the conductor for $\alpha\not=1/2$, where 
no exact solution is available. I obtained a result (\ref{eq39}) to leading order in $\Delta$. It is expected to hold
for sufficiently large $\varepsilon/\Delta_r$ (but with $\varepsilon$ still sufficiently smaller than the cut-off scale $\Lambda$),
and for times $t\ll1/\Delta_r$. 
Again there are damped coherent oscillations with angular frequency $\varepsilon$. 
The power law singularity $G_\sigma''(t)\sim t^{-1}$ found for $\alpha=1/2$
generalises to $G_\sigma''(t)\sim t^{-2\alpha}$ for $\alpha\not=1/2$. Both the exact expression for $G_\sigma''(t)$ at $\alpha=1/2$ 
and the perturbative expression at arbitrary
$\alpha$ therefore diverge as $t\to 0^+$. This is due to the severe shake-up
that the excited impurity causes in the Fermi sea of the conductor. The interpretation is confirmed by calculating
the response charge $Q_{ll'}$ for $\alpha<1/2$, where it is not ultra-violet divergent. This reveals a simple relation (\ref{eq41}) 
between $Q_{ll'}$ and the impurity decay rate $W(\varepsilon)$, that due to Fermi sea shake up, displays a Fermi edge singularity.
For $\alpha\geq1/2$, the Fermi sea shake up induced by the excited impurity is so severe that the response charge diverges
if the ultra-violet cut-off $\Lambda$ is sent to infinity. 
The divergence reflects a strong dependence of $Q_{ll'}$ on short length-scale $\sim1/\Lambda$
physics for $\alpha>1/2$. 

A possible line for future enquiry is the generalisation of the present work to situations where the electrons in the conductor
are described by a non-equilibrium distribution function.  
This is known to affect the impurity decay rate $W(\varepsilon)$ in
a non-trivial way\cite{muz03,aba05} and may therefore be expected to modify electron-electron correlations
in a similarly interesting manner. In order to study these correlations, one will have to solve the following technical problem.
Due to the difference in Fermi energy between left and right incident electrons in a voltage biased conductor,
channel space becomes entangled with the impurity state in such a way that
the diagonal representation of the Hamiltonian derived in Sec. \ref{s2} is of no use.\cite{muz}

\end{document}